\documentclass[pdflatex,sn-mathphys-num]{sn-jnl}

\usepackage{graphicx}%
\usepackage{multirow}%
\usepackage{amsmath,amssymb,amsfonts}%
\usepackage{amsthm}%
\usepackage{mathrsfs}%
\usepackage[title]{appendix}%
\usepackage{xcolor}%
\usepackage{textcomp}%
\usepackage{manyfoot}%
\usepackage{booktabs}%
\usepackage{algorithm}%
\usepackage{algorithmicx}%
\usepackage{algpseudocode}%
\usepackage{listings}%
\usepackage{graphicx}
\usepackage{slashed}
\usepackage{comment}

\theoremstyle{thmstyleone}%
\theoremstyle{thmstyletwo}%

\theoremstyle{thmstylethree}%

\begin{document}

\title[Article Title]{Addressing Standard Model Tensions via $X_{17}$ Vector Boson}


\author*[1,2]{\fnm{Raoul} \sur{Serao}}\email{rserao@unisa.it}

\author[3]{\fnm{Aniello} \sur{Quaranta}}\email{aniello.quaranta@unicam.it}
\equalcont{These authors contributed equally to this work.}

\author[1,2]{\fnm{Antonio} \sur{Capolupo}}\email{acapolupo@unisa.it}
\equalcont{These authors contributed equally to this work.}

\affil*[1]{\orgdiv{Department of Physics}, \orgname{University of Salerno}, \orgaddress{\street{Via Giovanni Paolo II, 132}, \city{Fisciano}, \postcode{84084}, \state{}, \country{Italy}}}

\affil[2]{\orgdiv{INFN}, \orgname{gruppo collegato di Salerno}, \orgaddress{\street{Via Giovanni Paolo II, 132}, \city{Fisciano}, \postcode{84084}, \state{}, \country{Italy}}}

\affil[3]{\orgdiv{School of Science and Technology}, \orgname{University of Camerino}, \orgaddress{\street{Via Madonna delle Carceri}, \city{Camerino}, \postcode{610101}, \state{}, \country{Italy}}}


\abstract{We investigate the effects of introducing a new vector boson on existing discrepancies within the Standard Model. Our analysis highlights the potential of this particle to alleviate these tensions while serving as a portal to the dark sector. This scenario provides a promising avenue for exploring extensions beyond the Standard Model and motivates further experimental and theoretical studies.}




\maketitle

\section{Introduction}\label{sec1}

The Standard Model (SM) of particle physics is among the most successful frameworks in modern science, providing an accurate description of fundamental particles and the electromagnetic, weak, and strong interactions. Its predictions have been extensively verified, most notably with the discovery of the Higgs boson at the LHC in 2012 \cite{higgs2012}. Nevertheless, the SM is generally regarded as an effective field theory, expected to break down at higher energy scales where new physics should appear. Several open questions drive the search for extensions beyond the SM (BSM).  
These include the origin and nature of dark matter and dark energy, particle mixing and the matter-antimatter asymmetry\cite{darkmatter1,dm1, darkmatter2,dm2,darkmatter3,dm3,darkmatter4,dm4,darkmatter5,darkmatter6,dm6,darkmatter7,dm7,darkmatter8}.

In addition to the previously mentioned open problems, there exists a series of persistent "tensions" between Standard Model (SM) predictions and experimental measurements, which may provide valuable hints of new physics. A well-known case is the anomalous magnetic moment of the muon \cite{Marciano2016, Cazzaniga2021}. Defining $a_\mu=(g_\mu-2)/2$, one finds a deviation between the SM prediction $a_\mu^{SM}$ and the experimental determination $a_\mu^{EXP}$. The latest results of the Muon g-2 collaboration \cite{Abi2021} confirm a discrepancy with respect to the SM expectation \cite{Patrignani2016,Aoyama2020}. Combining the Brookhaven \cite{Bennett2004} and Fermilab \cite{Abi2021} data yields a $4.2\sigma$ deviation, quantified as $\Delta a_\mu = (251 \pm 59)\times 10^{-11}$. A similar comparison can be made for the electron, where the extremely precise measurement $a^{EXP}_e=1159652180.73\times 10^{-12}$ leads to $\Delta a_e = (4.8 \pm 3.0)\times 10^{-13}$ \cite{Hannake,Nature2020}.

Another anomaly arises in the study of the Lamb shift in muonic atoms. For muonic hydrogen, the observed energy differences between the $2S$ and $2P$ states deviate from the SM predictions by $\delta E_\mu^H=(-0.363,-0.251)\,\mathrm{meV}$ \cite{muon1, muon2}.  

Further indications of possible BSM physics come from precision electroweak measurements. Recently, the CMS collaboration reported a value for the $W$ boson mass of $80360 \pm 9.9 \,\mathrm{MeV}$ \cite{W1}. Since at tree level $M_W = g\nu/2$, with $\nu=246$ GeV the Higgs vacuum expectation value and $g$ the weak isospin coupling, new particles could alter this relation through loop corrections, potentially accounting for the observed discrepancy.

Consequently, the search for BSM phenomena is a central aim of current research. In this context, the proposed $X_{17}$ particle has generated considerable interest \cite{X1}. A group at the Institute for Nuclear Research (ATOMKI) in Hungary reported anomalies in the angular correlations of $e^-e^+$ pairs from nuclear transitions \cite{X2,X3}. In particular, decays of an excited state of $^8$Be revealed an excess consistent with the emission of a neutral boson of about 17 MeV. A subsequent study of $^4$He nuclei confirmed a similar anomaly \cite{X4}, reinforcing the hypothesis of a new particle, informally dubbed “$X_{17}$” \cite{X5,X6}.  

Theoretical interpretations of $X_{17}$ vary, with proposals including a protophobic vector boson or a mediator between visible and dark matter. If confirmed, such a discovery would provide a new force carrier in the MeV regime and a possible connection to the dark sector. However, independent verification and further precision studies are required before firm conclusions can be drawn. The potential existence of $X_{17}$ would represent a major breakthrough in particle physics, reshaping our understanding of fundamental interactions and dark matter.
Based on the paper \cite{dm5}, we show that the introduction of the $X_{17}$ boson could explain some of the tensions between the Standard Model and the experimental results. In particular we focus on muon anomalous magnetic moment, Lamb shift and $W$ boson mass.

{\it Lepton magnetic moment}
 \begin{figure}[t]
 \centering
\includegraphics[width=0.7 \textwidth]{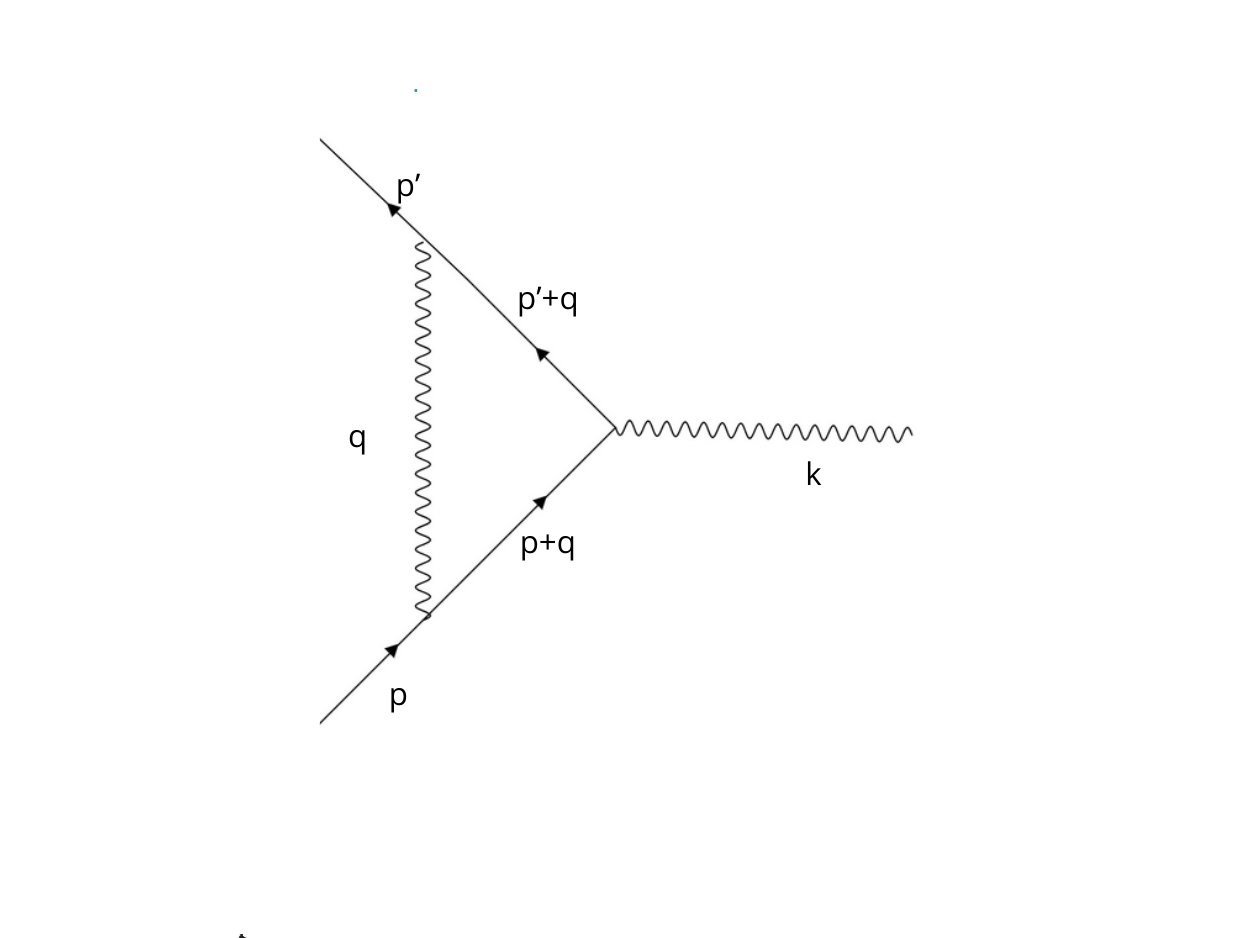}
\caption{Reference Feynman diagram for the calculation of the one loop correction to the magnetic dipole moment.}
\end{figure}

If this boson exist, it couples to charged fermion. This could give a correction to the $g$- factor.
The relevant diagram at one loop is that of Fig. 1 and the vertex correction is
\begin{eqnarray}\label{Vertex}
\nonumber &&\delta \Gamma^{\mu} (q) = \int \frac{d^4 q}{(2\pi)^4} \bar{u} (p') (-i \epsilon_l  \gamma^{\lambda})\frac{i(\slashed{p}'+\slashed{q}+m_l)}{(p'+q)^2-m_l^2}(e \gamma^\mu) \\ \nonumber
&&\times \frac{i(\slashed{p}+\slashed{q}+m_l)}{(p+q)^2-m_l^2} (-i \epsilon_l  \gamma^\nu)\frac{i}{q^2+M^2_X}\biggl(-g^{\nu\lambda}+\frac{q^\nu q^\lambda}{M^2_X}\biggr) u(p) ,   \\
\end{eqnarray}
where $\epsilon_l$ is the coupling constant  between the $X_{17}$ vector boson and lepton $l$.
The correction to the lepton $g$- factor is:
 \begin{equation}
 \label{deltaa}
    a_l^X= \frac{\alpha}{2\pi} \epsilon_l^2 \lambda_l \int_0^1 dx \frac{x^2(1-x)}{\lambda_lx^2-x+1}=\frac{\alpha}{2\pi} \epsilon_l^2 f(\lambda_l).
 \end{equation}
 where we have introduced the adimensional parameter $\lambda_l=m_l^2/M^2_X$ and $\epsilon_\mu$ is the coupling constant between $X_{17}$ and the muon.
 It is evident that, since the correction depends on the adimensional parameter $\lambda_l$, the introduction of $X_{17}$ could explain why the tension between the experimental results and theoretical predictions is greater for muon. Moreover we can expect that the tension for the tau is greater than the muon one.
By attributing these discrepancies to the corrections induced by the $X_{17}$ boson, it becomes possible to determine its coupling to the electron and the muon.
The relevant inequalities are:
\begin{eqnarray}\label{mome}
 \delta a_e=a_{e , EXP} - a_{e, SM} \leq 4.8 \times 10^{-13},
\end{eqnarray}
\begin{eqnarray}\label{momm}
 \delta a_\mu= a_{\mu , EXP} - a_{\mu, SM} \leq 2.51\times 10^{-9} \ ,
\end{eqnarray}
where we have used the latest available average values for the Standard Model prediction and experimental results \cite{Abi2021,Patrignani2016,Aoyama2020,Bennett2004}.
The upper bound on the coupling of the $X_{17}$ to the muon is $|\epsilon_\mu| <2.154 \times 10^{-4}$, and to the electron $|\epsilon_e|< 1.02 \times 10^{-4}$.

{\it Lamb shift}
The presence of this new vector boson could affects also the Lamb shift.
The nonrelativistic potential between the proton and muon due to the exchange of a vector boson is:
\begin{equation}
\label{Pot}
    V_X(r)=\frac{\epsilon_\mu \epsilon_p}{e^2} \frac{\alpha e^{-M_X r}}{r},
\end{equation}
This potential gives an additional contribution to the Lamb shift in the $2S_{1/2}-2P_{3/2}$ transition:
\begin{equation}
\label{deltaH}
\begin{split}
    \delta E^H_X&=\int dr r^2V_X(r)(\|R_{20}(r)\|^2-\|R_{21}\|^2)\\
    &= \frac{\alpha}{2 a_H^3} \biggl(\frac{\epsilon_\mu \epsilon_p}{e^2}\biggr) \frac{f(a_H M_X)}{M_X^2} \ .
    \end{split}
\end{equation}
where the function $f(x)$ is $f(x)=\frac{x^4}{(1+x)^4}$ and $a_{H}=(\alpha m_{\mu p})^{-1}$ is the Bohr radius of the system, $m_{\mu p}$ is the reduced mass of the muonic hydrogen and $R_{nl}$ are the radial wave function.
Considering the latest experimental value for $\delta E^{H}_{\mu}$ \cite{muon1,muon2} we can derive $\epsilon_p$.
This yields an upper bound on the (modulus of the) coupling between the $X_{17}$ and the proton in terms of the $X_{17}$ mass $M_X$ and of the Lamb shift deviation $\delta E^{H}_{\mu}$, as depicted in Fig. 2.
Considering a range of values for $M_{X}$ between $[16.7, 17.2] \ \mathrm{MeV}$, and for $\delta E_\mu^H$ between $[-0.363, -0.251]\ \mathrm{meV}$, the range of values for the lower bound on the coupling $\epsilon_p$ will be between $[-0.04260,-0.02982]$. Consequently the upper bound on $|\epsilon_p|$ ranges in $[0.02982,0.04260]$.
We notice that there is a sign ambiguity on $\epsilon_{\mu}$: the values of Fig. 2 are obtained by assuming a positive muon coupling $\epsilon_{\mu} > 0$, which leads to a negative proton coupling $\epsilon_{p} < 0$.
\begin{figure}[t]
\centering
\includegraphics[width= 0.6 \textwidth]{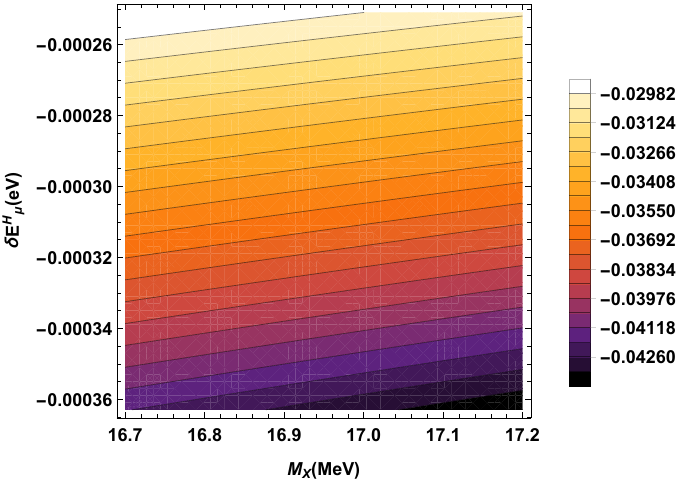}
\caption{ (Color online) Contour plot of the lower bound for the coupling constant $\epsilon_p$ between the $X_{17}$ and the proton as a function of the mass of $X_{17}$ and the experimental data on muonic hydrogen Lamb shift  $\delta E_\mu^H$ \cite{muon1, muon2}, as obtained from equation \eqref{deltaH}. The legend on the side shows the values of the coupling constant $\epsilon_p$ for different values of $M_X$ and $\delta E_\mu^H$. We consider $\epsilon_{\mu} \simeq 2.154 \times 10^{-4}$ .}. 
\end{figure}

{\it W boson mass}
In this section we consider how the kinetic miixing between the hypercharge boson $b$ and the dark photon could generate a shift on the $W$ mass, following the method described in \cite{W00}.
The shift on the $W$ mass is given by
\begin{equation}\label{Wmassshift}
    \Delta M_W=M_W-M_{W,SM}=-\frac{M_{W,SM}s_W^2\xi^2}{2(c_W^2-s_W^2)^2(1-r^2)},
\end{equation}
 where $c_W=\cos{\theta_W}$,$s_W=\cos{\theta_W}$, $\theta_W$ is the Weinberg angle, $\xi$ is the kinetic mixing parameter and  $r=\frac{M_Z}{M_X}$.
By setting $\Delta M_W\leq 10 \ \mathrm{MeV}$, corresponding to the uncertainty from the latest measurements \cite{W1}, we can derive an upper bound on the kinetic mixing parameter: $|\xi|<2.2 \times 10^{-2}$. 

{\it Conclusion}
In this study, we have investigated how the introduction of a new vector boson can influence the observed discrepancies within the Standard Model, highlighting its potential to alleviate existing tensions in precision measurements. Our analysis suggests that such a mediator could not only account for part of these deviations but also serve as a plausible connection to the dark sector. By acting as a portal between ordinary matter and dark matter, this particle may provide new guidance on viable extensions of the Standard Model. Future studies, combining improved experimental data and refined theoretical calculations, will be crucial to further test this scenario and delineate its implications for particle physics beyond the Standard Model.

\backmatter

\bmhead{Acknowledgements}

We acknowledge partial financial support from MUR and INFN, A.C. also acknowledges the COST Action CA1511 Cosmology
and Astrophysics Network for Theoretical Advances and Training
Actions (CANTATA).

\bibliography{sn-bibliography}

\end{document}